\documentclass[a4paper,11pt]{article}
\usepackage{pos}
\usepackage{graphicx}
\usepackage{siunitx}
\usepackage{upgreek}
\usepackage[capitalize]{cleveref}
\usepackage{comment}
\usepackage{enumitem}
\usepackage{wrapfig}
\usepackage{microtype}

\DeclareSIUnit{\litre}{l}

\title{Monitoring and performance of AugerPrime}
\ShortTitle{Monitoring and performance of AugerPrime}

\author*[a]{Belén Andrada}
\onbehalf{for the Pierre Auger Collaboration$^b$}

\affiliation[a]{Instituto de Tecnología y Detección en Astropartículas (CNEA, CONICET, UNSAM), Buenos Aires, Argentina}

\affiliation[b]{Observatorio Pierre Auger, Av.\ San Mart{\'\i}n Norte 304, 5613 Malarg\"ue, Argentina\\
Full author list: {\rm\url{https://www.auger.org/archive/authors_icrc_2025.html}}}

\emailAdd{spokespersons@auger.org}

\abstract{The upgrade of Pierre Auger Observatory, AugerPrime, is a multi-hybrid system designed to improve the sensitivity and precision of ultra-high-energy cosmic ray measurements. 
It includes scintillator detectors positioned both atop the enhanced Water-Cherenkov detectors and buried nearby for direct muon measurements, along with radio and fluorescence detectors. 
In this contribution, we present an overview of the monitoring tools developed for all the components of AugerPrime, focusing on real-time performance assessment and long-term stability metrics. 
By continuously tracking key parameters, we can identify potential issues early, enabling timely interventions and improving overall data quality. 
These strategies are crucial for maintaining the long-term reliability of the measurements taken at the Auger Observatory and providing high-quality data for cosmic ray research in the coming decades.}

\ConferenceLogo{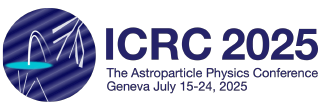}

\FullConference{39th International Cosmic Ray Conference (ICRC2025)\\
 15–24 July 2025\\
Geneva, Switzerland\\}

\begin{document}
\maketitle

\section{The AugerPrime Surface Detector}

The study of ultra-high-energy cosmic rays (UHECRs), with energies exceeding $10^{17}$\,eV, remains one of the most compelling open questions in astroparticle physics. Key challenges include understanding their origin, acceleration mechanisms, and mass composition. The Pierre Auger Observatory was designed to address these questions by combining a large surface detector array with fluorescence telescopes, enabling the hybrid observation of extensive air showers over a wide energy range.

\begin{figure}
    \centering
    \includegraphics[width=\linewidth]{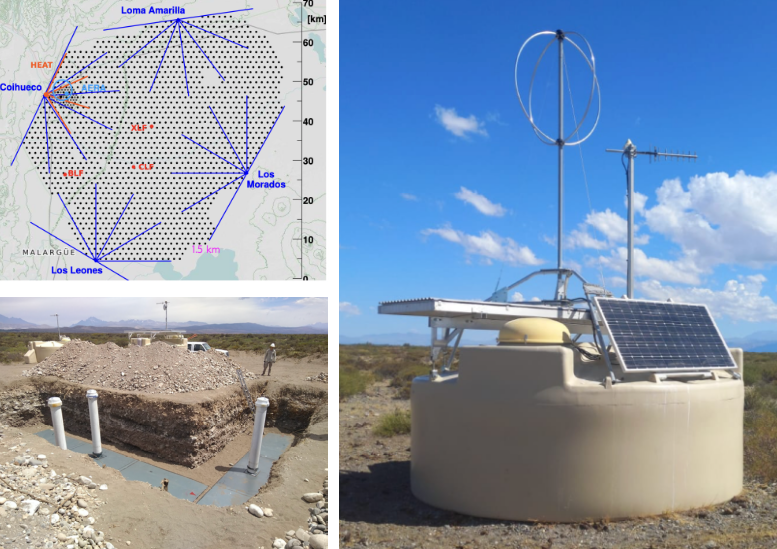}
    \caption{Top-left: layout of the Pierre Auger Observatory. Black dots represent surface detector stations and blue lines indicate the approximate field of view of the fluorescence telescopes in its four sites. The SD-750 and SD-433 arrays are located near the Coihueco site. Bottom-left: Underground Muon Detector station under deployment. Right: upgraded Water-Cherenkov detector, with a Scintillator detector and a Radio antenna. }
    \label{fig:detectors}
\end{figure}

The Pierre Auger Observatory is located in Malargüe, Argentina, and comprises over 1660 surface detector stations arranged in triangular grids with spacings of \SI{1500}{\meter}, \SI{750}{\meter}, \SI{433}{\meter} (called SD-1500, SD-750, and SD-433 respectively) and, covering a total area of approximately \SI{3000}{\square\kilo\meter}~\cite{auger_nim}. The surface array is complemented by 27 fluorescence telescopes installed at four peripheral buildings, as can be seen in the top-left panel of \cref{fig:detectors}. While the denser sub-arrays enable low-energy event reconstruction, the fluorescence detector overlooks the atmosphere above the surface detector area to observe the longitudinal development of air showers and provides both mass-sensitive observables and a model-independent energy reconstruction.

During its first phase of operation (Phase I), the Observatory recorded high-quality data that led to significant advances, including precise measurements of the UHECR energy spectrum, constraints on photon and neutrino fluxes, and composition-sensitive observables~\cite{uhecr24_overview}. These results suggest that the primary mass composition varies with energy while highlighting persistent uncertainties in hadronic interaction models at the highest energies. 

To address these challenges and enhance composition sensitivity, the Auger Observatory has undergone a major upgrade known as AugerPrime~\cite{auger_prime}. As part of this upgrade, each Water-Cherenkov Detector (WCD) that is part of the surface array has been equipped with a small photomultiplier tube (SPMT) to extend its dynamic range and improve signal reconstruction near the shower core. In addition, the original electronics have been replaced by the Upgraded Unified Board (UUB), a new data acquisition system that digitizes signals from all detectors at \SI{120}{\mega\hertz} with 12-bit resolution~\cite{auger_uub}. The UUB provides improved timing, triggering, and calibration functionality, ensuring compatibility with Phase I data while supporting the new AugerPrime instrumentation.

In addition to the WCD enhancements, each surface station now includes a Surface-Scintillator Detector (SSD) mounted above the WCD and a Radio Detector (RD) station installed on a mast, as can be seen in the right panel of \cref{fig:detectors}. A dedicated Underground Muon Detector (UMD), almost fully deployed in the SD-750 and SD-433 regions (see bottom left panel of \cref{fig:detectors}), complements these measurements by providing a direct view of the muonic component. Together, these additions allow for multi-component air shower reconstruction with improved sensitivity to mass composition.

A critical component of AugerPrime is the implementation of comprehensive monitoring systems across all detectors. These systems enable real-time tracking of key performance parameters, identification of potential issues, and long-term assessments of detector stability. In this work, we describe the monitoring infrastructure developed for each AugerPrime component and assess its performance during the initial years of Phase II data taking. 

\section{Water-Cherenkov Detectors}\label{sec:WCD}

The Water-Cherenkov Detectors form the core of the Surface Detector array and have operated reliably since the beginning of Phase I. Each WCD consists of a cylindrical tank filled with \SI{12000}{\litre} of purified water, instrumented with three downward-facing large photomultiplier tubes (LPMTs) mounted at the top. These PMTs detect Cherenkov light from charged particles and provide signals over a wide dynamic range suitable for most shower geometries.

As already mentioned, as part of the AugerPrime upgrade, a fourth, small PMT was added to extend the dynamic range, and the original electronics were replaced by the Upgraded Unified Board. The SPMT allows accurate measurement of high particle densities near the shower axis, while the UUB provides enhanced digitization, timing, and calibration capabilities~\cite{auger_prime}. This extends the WCD dynamic range by more than an order of magnitude, from a few hundred VEM (vertical equivalent muon) with the large PMTs to nearly 20,000\,VEM with the SPMT.

\begin{figure}
    \centering
    \includegraphics[width=\textwidth]{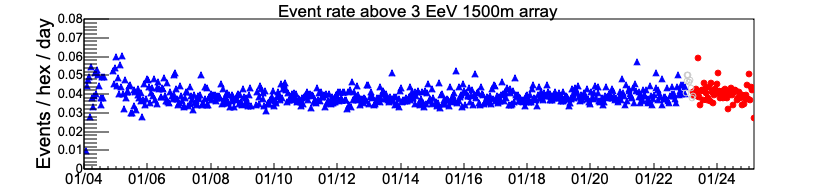}
  \caption{Daily rate of high-quality events per active hexagon for the SD-1500. The energy threshold ensures full trigger efficiency. Blue triangles indicate Phase I data, and red circles correspond to Phase II data, while the open gray circles show the transition period.}
  \label{fig:wcd_eventrate}
\end{figure}

The stability of the WCD array is shown in \cref{fig:wcd_eventrate}, which presents the daily rate of high-quality events per active hexagon for the SD-1500, as an example. The indicated energy threshold is chosen to ensure full trigger efficiency. Blue triangles indicate Phase I data, and red circles correspond to Phase II data, while the open gray circles show the transition period between the two. Rates remain stable over time, demonstrating consistent array performance during and after the deployment.

The SPMT, a 1-inch Hamamatsu R8619, cannot be directly calibrated with atmospheric muons due to its small photocathode area. A cross-calibration procedure is used instead, converting the integrated ADC signal to VEM via a linear relation, $S_{\text{VEM}} = \beta\,Q_{\text{ADC counts}}$, with a conversion factor $\beta$ determined to better than $2.5\,\%$~\cite{spmt_proceeding}.

\begin{figure}
    \centering
    \includegraphics[width=\linewidth]{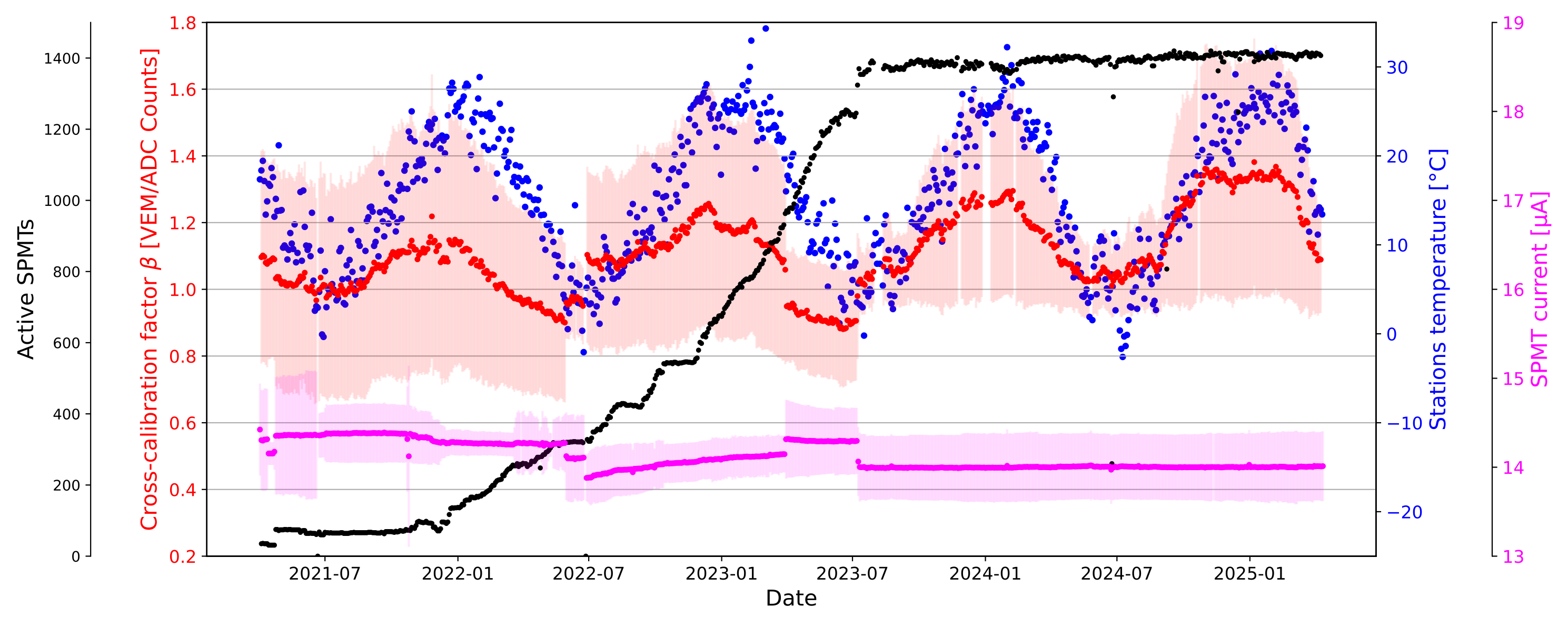}
    \caption{Evolution of SPMT-related parameters: number of active SPMTs (black), calibration factor $\beta$ (red), station temperature (blue), and SPMT current (magenta), from 2021 to 2025.}
    \label{fig:spmt_evolution}
\end{figure}

The long-term evolution of SPMT-related parameters can be seen in \cref{fig:spmt_evolution}. The number of active SPMTs (black) increases until mid-2023 as deployment progresses. The calibration factor $\beta$ (red) shows a seasonal modulation of 8–10\,$\%$, consistent with the temperature dependence of SPMT gain. Temperature (blue), taken from sensors on the LPMTs, and SPMT current (magenta) reflect environmental trends and their effect on the hardware. In particular, the current stabilizes in mid-2023 following the activation of automatic high-voltage regulation.

The UUB digitizes waveforms from all detectors and provides unified acquisition across the upgraded Surface Detector~\cite{uub_proceeding}. It also maintains backward compatibility with Phase I triggers to support hybrid operation. Monitoring of voltage rails, temperature, and acquisition rates enables early issue detection and long-term stability tracking.

\section{Surface Scintillator Detectors}\label{sec:SSD}

Each Surface Scintillator Detector consists of two segmented plastic scintillator planes mounted above the WCD. Wavelength-shifting fibers collect scintillation light and guide it to a 1.5-inch Hamamatsu R9420 PMT. Unlike WCDs, which respond to both electromagnetic and muonic components, the SSDs are more sensitive to the electromagnetic component, enabling improved composition discrimination.

Calibration is primarily based on the response to atmospheric muons, which deposit a characteristic energy in the scintillator. This defines the minimum ionizing particle (MIP) signal, expressed in ADC counts, used as the reference unit for all signals~\cite{ssd_calibration}.

\begin{figure}
    \centering
    \includegraphics[width=\linewidth]{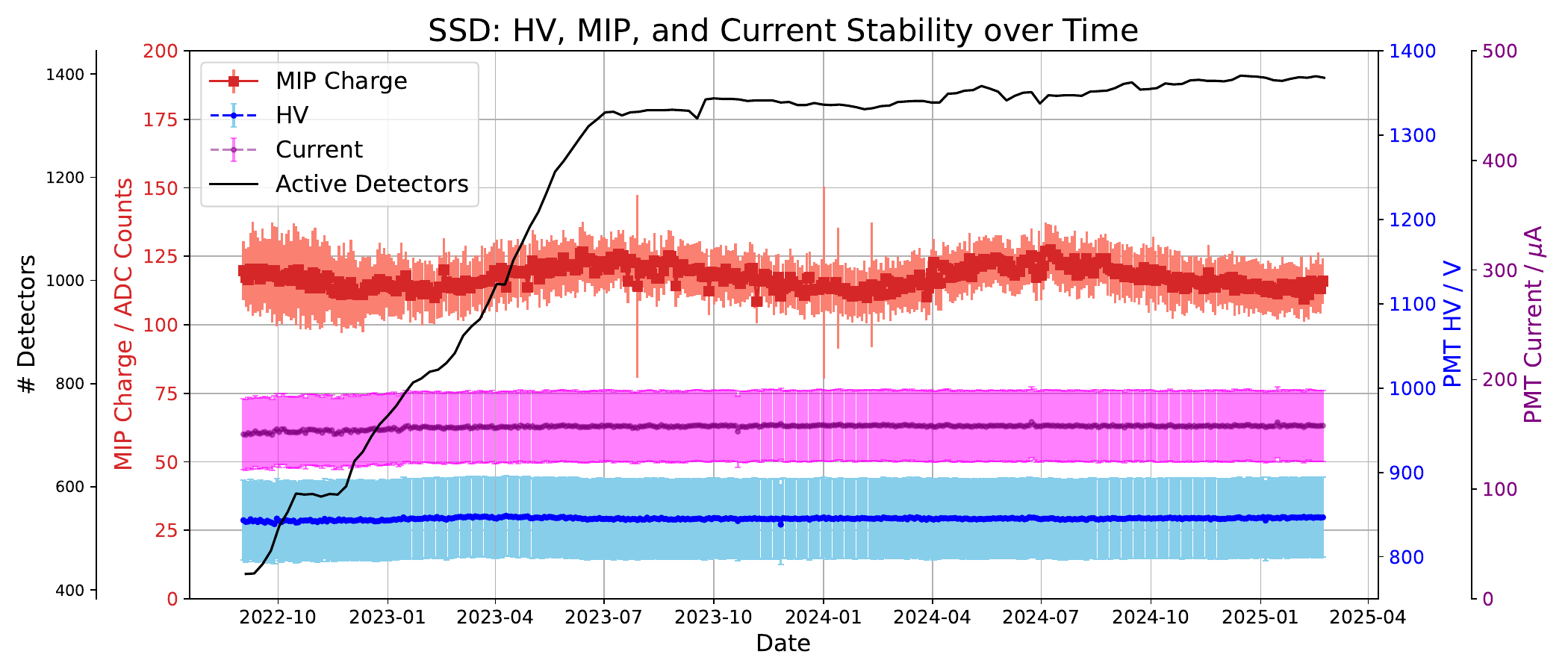}
    \caption{Daily averages of SSD MIP charge (red), PMT high voltage (blue), anode current (magenta), and the number of operational SSDs (black), grouped in 50-day intervals.}
    \label{fig:ssd-perf}
\end{figure}

The evolution of key monitoring parameters from late 2022 to early 2025 is shown in \cref{fig:ssd-perf}: MIP charge (red), PMT high voltage (blue), and anode current (magenta), along with the number of operational SSDs (black), averaged in 50-day intervals. The number of active detectors increases steadily during 2023 and stabilizes in mid-year, following installation completion and the start of Phase II data taking. High voltage and current remain stable, with no clear seasonal trend. In contrast, the MIP charge shows a 4–5\,$\%$ seasonal modulation and diurnal variation (not shown here), attributed to the temperature dependence of PMT gain. Electronics temperatures remain stable, suggesting the gain is the primary source of variation~\cite{ssd_performance}.

\section{Radio Detector}\label{sec:RD}

The Radio Detector measures coherent radio emission from extensive air showers in the 30–\SI{80}{\mega\hertz} band. Each RD station includes a dual-polarized Short Aperiodic Loaded Loop Antenna (SALLA) mounted above the WCD (see right panel of \cref{fig:detectors}). The two orthogonal channels are aligned relative to the local geomagnetic field to maximize sensitivity to the dominant emission mechanisms. Signals are amplified, filtered, and digitized at \SI{250}{\mega\hertz} with 12-bit resolution and are read out upon triggering by the associated WCD, ensuring precise timing across components. 

\begin{figure}
      \centering
      \includegraphics[width=\linewidth]{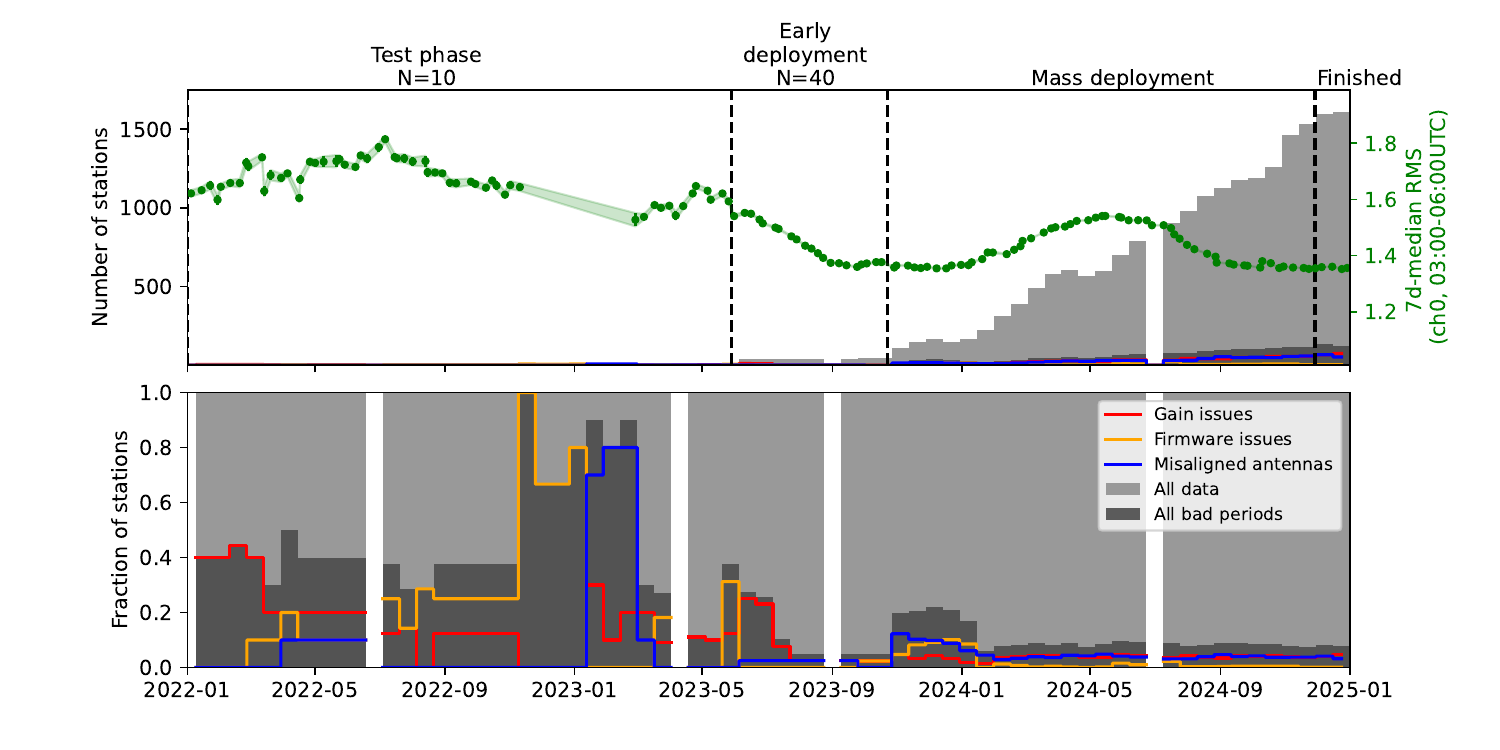}
      \caption{Number of operational RD stations from 2022 to 2025, with flagged periods (gray) marking quality issues. Vertical dashed lines indicate deployment milestones. Green markers in the top panel show the 7-day median RMS voltage, averaged over all active stations, calculated using data from 03:00–06:00\,UTC only.}
      \label{fig:RD_MOLTP}
\end{figure}

The number of operational RD stations over time can be observed in \cref{fig:RD_MOLTP} as well as flagged periods where quality issues were detected. These include firmware errors, antenna misalignments, or gain instability, all identified through automated monitoring (see also~\cite{ref:ICRC2025RDFirstData}). The vertical lines mark transitions between deployment phases. Deployment began with a small number of stations in 2022, followed by mass installation between November 2023 and November 2024. By the end of 2024, nearly all stations reached stable performance.

An additional long-term observable is the average RMS voltage measured at each station, representing the total noise level in the signal band. About half of this RMS is due to diffuse galactic background emission, which is used as a calibration reference. The RMS increases during thunderstorms or transient events and is otherwise stable. Narrow-band radio-frequency interference (RFI) is filtered prior to RMS estimation to isolate broadband background variations~\cite{ref:ARENA2021Tomas}. To ensure consistent sky coverage throughout the year, RMS values shown in green on the top panel of \cref{fig:RD_MOLTP} are computed only from data recorded between 03:00 and 06:00\,UTC each day. This fixed time window reveals the seasonal modulation due to the galactic signal as the galaxy drifts across the night sky.

\section{Underground Muon Detector}\label{sec:UMD}

The Underground Muon Detector is installed in the SD-750 and SD-433 arrays to provide a direct measurement of the muonic component of extensive air showers. Each station includes three scintillator modules composed of 64 plastic scintillator strips, buried at a depth of \SI{2.3}{\meter} (see bottom right panel of \cref{fig:detectors}) and coupled to silicon photomultiplier (SiPM) arrays. Each SiPM detects light produced in a single strip via a wavelength-shifting fiber, enabling highly segmented detection of individual muons.

The UMD operates in binary mode~\cite{uhecr24_umd}, where each strip is read out at \SI{320}{\mega\hertz} and a digital ``1'' (one) is recorded whenever the signal exceeds a predefined threshold. This approach provides compact data with precise time resolution and robust noise suppression. Muon candidates are identified using a muon-like pattern, defined as at least four consecutive ``1'' bits in the binary trace (i.e., a pattern such as 1111) in the digitized trace. This pattern reflects the typical time structure of a muon signal and is used to count activated strips per event.

\begin{figure}
    \centering
   \includegraphics[width=\linewidth]{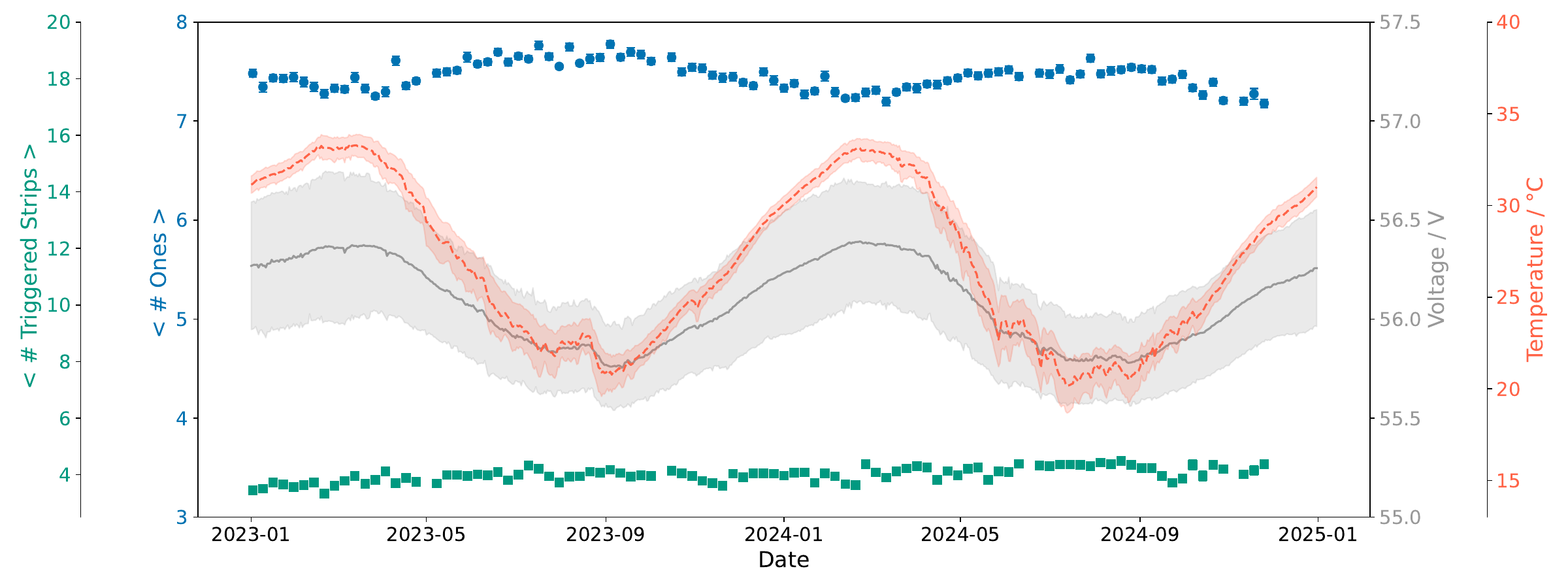}
    \caption{Daily averages of UMD monitoring parameters across all operational modules from 2023 to 2025: temperature measured at the high-voltage source (pink dashed line), applied voltage (gray solid line), average number of ones per trace (blue circles), and the average number of triggered strips per event with muon-like pattern (green squares).}
    \label{fig:umd-aging}
\end{figure}

The long-term evolution of key monitoring observables for the UMD is presented in \cref{fig:umd-aging}. These include the temperature measured at the high-voltage (HV) source (red dashed line), the applied HV (gray solid line), the average number of ones per trace (blue circles), and the average number of activated strips with a muon-like pattern per event (green squares). All variables are averaged over all operational modules in the array. A seasonal trend is observed in both temperature and voltage, as expected from the environmental modulation at the site. The HV source includes an internal temperature sensor and automatically adjusts the applied voltage to compensate for temperature-dependent shifts in the SiPM breakdown voltage. As the temperature decreases, the breakdown voltage of the SiPMs drops, increasing the overvoltage and, consequently, the gain. This results in larger signal amplitudes, effectively lowering the relative threshold and increasing the number of samples above the threshold. The resulting anticorrelation between temperature and the number of ones is clearly visible.

Despite the variations in gain and trace width, the average number of triggered strips per event remains stable over time. This confirms that the muon pattern recognition and event-level response are not significantly affected by temperature fluctuations or gain drift, highlighting the robustness of the binary mode and the effectiveness of the temperature compensation system. A gradual decrease in the number of ones is observed, corresponding to an average reduction of about 0.7\,$\%$ per year~\cite{icrc23_umd}. This effect may reflect aging in the scintillators, SiPMs, or associated electronics and will continue to be monitored in future stability studies.

\section{Conclusions}

The AugerPrime upgrade has been successfully implemented across the Surface Detector array, with all components now integrated into a unified and modern data acquisition system. The deployment of new instrumentation— including SPMTs, SSDs, RDs, and UMDs—alongside the Upgraded Unified Board has significantly expanded the detector capabilities and maintained full compatibility with Phase I data.

In this contribution, we presented the monitoring infrastructure developed for each component and assessed its performance during the initial years of Phase II. The Water-Cherenkov Detectors exhibit stable operation and extended dynamic range. The Surface Scintillator Detectors maintain consistent calibration with well-understood seasonal variations. The Radio Detectors demonstrate reliable deployment and continuous background monitoring. The Underground Muon Detectors show stable response in binary mode, with robust muon pattern recognition and only minor aging effects observed.

Continuous monitoring of calibration parameters, environmental dependencies, and detector uptime is essential for ensuring high-quality data over the long term. These tools are key to maintaining the stability of the upgraded array and will support future performance studies and physics analyses throughout Phase II.

\clearpage
\section*{The Pierre Auger Collaboration}

{\footnotesize\setlength{\baselineskip}{10pt}
\noindent
\begin{wrapfigure}[11]{l}{0.12\linewidth}
\vspace{-4pt}
\includegraphics[width=0.98\linewidth]{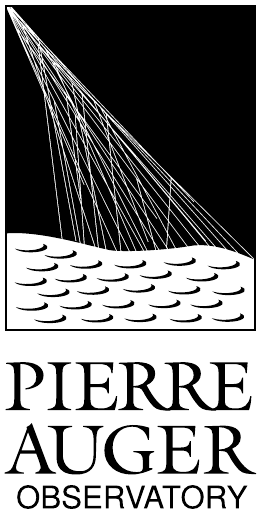}
\end{wrapfigure}
\begin{sloppypar}\noindent
A.~Abdul Halim$^{13}$,
P.~Abreu$^{70}$,
M.~Aglietta$^{53,51}$,
I.~Allekotte$^{1}$,
K.~Almeida Cheminant$^{78,77}$,
A.~Almela$^{7,12}$,
R.~Aloisio$^{44,45}$,
J.~Alvarez-Mu\~niz$^{76}$,
A.~Ambrosone$^{44}$,
J.~Ammerman Yebra$^{76}$
% ...

\end{sloppypar}

\begin{center}
\rule{0.1\columnwidth}{0.5pt}
\raisebox{-0.4ex}{\scriptsize$\bullet$}
\rule{0.1\columnwidth}{0.5pt}
\end{center}

\newpage
\vspace{1ex}
\begin{description}[labelsep=0.2em,align=right,labelwidth=0.7em,labelindent=0em,leftmargin=2em,noitemsep,before={\renewcommand\makelabel[1]{##1 }}]
\item[$^{1}$] Centro At\'omico Bariloche and Instituto Balseiro (CNEA-UNCuyo-CONICET), San Carlos de Bariloche, Argentina
\item[$^{2}$] Departamento de F\'\i{}sica and Departamento de Ciencias de la Atm\'osfera y los Oc\'eanos, FCEyN, Universidad de Buenos Aires and CONICET, Buenos Aires, Argentina
\item[$^{3}$] IFLP, Universidad Nacional de La Plata and CONICET, La Plata, Argentina
\item[$^{4}$] Instituto de Astronom\'\i{}a y F\'\i{}sica del Espacio (IAFE, CONICET-UBA), Buenos Aires, Argentina
\item[$^{5}$] Instituto de F\'\i{}sica de Rosario (IFIR) -- CONICET/U.N.R.\ and Facultad de Ciencias Bioqu\'\i{}micas y Farmac\'euticas U.N.R., Rosario, Argentina
\item[$^{6}$] Instituto de Tecnolog\'\i{}as en Detecci\'on y Astropart\'\i{}culas (CNEA, CONICET, UNSAM), and Universidad Tecnol\'ogica Nacional -- Facultad Regional Mendoza (CONICET/CNEA), Mendoza, Argentina
\item[$^{7}$] Instituto de Tecnolog\'\i{}as en Detecci\'on y Astropart\'\i{}culas (CNEA, CONICET, UNSAM), Buenos Aires, Argentina
\item[$^{8}$] International Center of Advanced Studies and Instituto de Ciencias F\'\i{}sicas, ECyT-UNSAM and CONICET, Campus Miguelete -- San Mart\'\i{}n, Buenos Aires, Argentina
\item[$^{9}$] Laboratorio Atm\'osfera -- Departamento de Investigaciones en L\'aseres y sus Aplicaciones -- UNIDEF (CITEDEF-CONICET), Argentina
\item[$^{10}$] Observatorio Pierre Auger, Malarg\"ue, Argentina
\item[$^{11}$] Observatorio Pierre Auger and Comisi\'on Nacional de Energ\'\i{}a At\'omica, Malarg\"ue, Argentina
\item[$^{12}$] Universidad Tecnol\'ogica Nacional -- Facultad Regional Buenos Aires, Buenos Aires, Argentina
\item[$^{13}$] University of Adelaide, Adelaide, S.A., Australia
\item[$^{14}$] Universit\'e Libre de Bruxelles (ULB), Brussels, Belgium
\item[$^{15}$] Vrije Universiteit Brussels, Brussels, Belgium
\item[$^{16}$] Centro Brasileiro de Pesquisas Fisicas, Rio de Janeiro, RJ, Brazil
\item[$^{17}$] Centro Federal de Educa\c{c}\~ao Tecnol\'ogica Celso Suckow da Fonseca, Petropolis, Brazil
\item[$^{18}$] Instituto Federal de Educa\c{c}\~ao, Ci\^encia e Tecnologia do Rio de Janeiro (IFRJ), Brazil
\item[$^{19}$] Universidade de S\~ao Paulo, Escola de Engenharia de Lorena, Lorena, SP, Brazil
\item[$^{20}$] Universidade de S\~ao Paulo, Instituto de F\'\i{}sica de S\~ao Carlos, S\~ao Carlos, SP, Brazil
\item[$^{21}$] Universidade de S\~ao Paulo, Instituto de F\'\i{}sica, S\~ao Paulo, SP, Brazil
\item[$^{22}$] Universidade Estadual de Campinas (UNICAMP), IFGW, Campinas, SP, Brazil
\item[$^{23}$] Universidade Estadual de Feira de Santana, Feira de Santana, Brazil
\item[$^{24}$] Universidade Federal de Campina Grande, Centro de Ciencias e Tecnologia, Campina Grande, Brazil
\item[$^{25}$] Universidade Federal do ABC, Santo Andr\'e, SP, Brazil
\item[$^{26}$] Universidade Federal do Paran\'a, Setor Palotina, Palotina, Brazil
\item[$^{27}$] Universidade Federal do Rio de Janeiro, Instituto de F\'\i{}sica, Rio de Janeiro, RJ, Brazil
\item[$^{28}$] Universidad de Medell\'\i{}n, Medell\'\i{}n, Colombia
\item[$^{29}$] Universidad Industrial de Santander, Bucaramanga, Colombia
\item[$^{30}$] Charles University, Faculty of Mathematics and Physics, Institute of Particle and Nuclear Physics, Prague, Czech Republic
\item[$^{31}$] Institute of Physics of the Czech Academy of Sciences, Prague, Czech Republic
\item[$^{32}$] Palacky University, Olomouc, Czech Republic
\item[$^{33}$] CNRS/IN2P3, IJCLab, Universit\'e Paris-Saclay, Orsay, France
\item[$^{34}$] Laboratoire de Physique Nucl\'eaire et de Hautes Energies (LPNHE), Sorbonne Universit\'e, Universit\'e de Paris, CNRS-IN2P3, Paris, France
\item[$^{35}$] Univ.\ Grenoble Alpes, CNRS, Grenoble Institute of Engineering Univ.\ Grenoble Alpes, LPSC-IN2P3, 38000 Grenoble, France
\item[$^{36}$] Universit\'e Paris-Saclay, CNRS/IN2P3, IJCLab, Orsay, France
\item[$^{37}$] Bergische Universit\"at Wuppertal, Department of Physics, Wuppertal, Germany
\item[$^{38}$] Karlsruhe Institute of Technology (KIT), Institute for Experimental Particle Physics, Karlsruhe, Germany
\item[$^{39}$] Karlsruhe Institute of Technology (KIT), Institut f\"ur Prozessdatenverarbeitung und Elektronik, Karlsruhe, Germany
\item[$^{40}$] Karlsruhe Institute of Technology (KIT), Institute for Astroparticle Physics, Karlsruhe, Germany
\item[$^{41}$] RWTH Aachen University, III.\ Physikalisches Institut A, Aachen, Germany
\item[$^{42}$] Universit\"at Hamburg, II.\ Institut f\"ur Theoretische Physik, Hamburg, Germany
\item[$^{43}$] Universit\"at Siegen, Department Physik -- Experimentelle Teilchenphysik, Siegen, Germany
\item[$^{44}$] Gran Sasso Science Institute, L'Aquila, Italy
\item[$^{45}$] INFN Laboratori Nazionali del Gran Sasso, Assergi (L'Aquila), Italy
\item[$^{46}$] INFN, Sezione di Catania, Catania, Italy
\item[$^{47}$] INFN, Sezione di Lecce, Lecce, Italy
\item[$^{48}$] INFN, Sezione di Milano, Milano, Italy
\item[$^{49}$] INFN, Sezione di Napoli, Napoli, Italy
\item[$^{50}$] INFN, Sezione di Roma ``Tor Vergata'', Roma, Italy
\item[$^{51}$] INFN, Sezione di Torino, Torino, Italy
\item[$^{52}$] Istituto di Astrofisica Spaziale e Fisica Cosmica di Palermo (INAF), Palermo, Italy
\item[$^{53}$] Osservatorio Astrofisico di Torino (INAF), Torino, Italy
\item[$^{54}$] Politecnico di Milano, Dipartimento di Scienze e Tecnologie Aerospaziali , Milano, Italy
\item[$^{55}$] Universit\`a del Salento, Dipartimento di Matematica e Fisica ``E.\ De Giorgi'', Lecce, Italy
\item[$^{56}$] Universit\`a dell'Aquila, Dipartimento di Scienze Fisiche e Chimiche, L'Aquila, Italy
\item[$^{57}$] Universit\`a di Catania, Dipartimento di Fisica e Astronomia ``Ettore Majorana``, Catania, Italy
\item[$^{58}$] Universit\`a di Milano, Dipartimento di Fisica, Milano, Italy
\item[$^{59}$] Universit\`a di Napoli ``Federico II'', Dipartimento di Fisica ``Ettore Pancini'', Napoli, Italy
\item[$^{60}$] Universit\`a di Palermo, Dipartimento di Fisica e Chimica ''E.\ Segr\`e'', Palermo, Italy
\item[$^{61}$] Universit\`a di Roma ``Tor Vergata'', Dipartimento di Fisica, Roma, Italy
\item[$^{62}$] Universit\`a Torino, Dipartimento di Fisica, Torino, Italy
\item[$^{63}$] Benem\'erita Universidad Aut\'onoma de Puebla, Puebla, M\'exico
\item[$^{64}$] Unidad Profesional Interdisciplinaria en Ingenier\'\i{}a y Tecnolog\'\i{}as Avanzadas del Instituto Polit\'ecnico Nacional (UPIITA-IPN), M\'exico, D.F., M\'exico
\item[$^{65}$] Universidad Aut\'onoma de Chiapas, Tuxtla Guti\'errez, Chiapas, M\'exico
\item[$^{66}$] Universidad Michoacana de San Nicol\'as de Hidalgo, Morelia, Michoac\'an, M\'exico
\item[$^{67}$] Universidad Nacional Aut\'onoma de M\'exico, M\'exico, D.F., M\'exico
\item[$^{68}$] Institute of Nuclear Physics PAN, Krakow, Poland
\item[$^{69}$] University of \L{}\'od\'z, Faculty of High-Energy Astrophysics,\L{}\'od\'z, Poland
\item[$^{70}$] Laborat\'orio de Instrumenta\c{c}\~ao e F\'\i{}sica Experimental de Part\'\i{}culas -- LIP and Instituto Superior T\'ecnico -- IST, Universidade de Lisboa -- UL, Lisboa, Portugal
\item[$^{71}$] ``Horia Hulubei'' National Institute for Physics and Nuclear Engineering, Bucharest-Magurele, Romania
\item[$^{72}$] Institute of Space Science, Bucharest-Magurele, Romania
\item[$^{73}$] Center for Astrophysics and Cosmology (CAC), University of Nova Gorica, Nova Gorica, Slovenia
\item[$^{74}$] Experimental Particle Physics Department, J.\ Stefan Institute, Ljubljana, Slovenia
\item[$^{75}$] Universidad de Granada and C.A.F.P.E., Granada, Spain
\item[$^{76}$] Instituto Galego de F\'\i{}sica de Altas Enerx\'\i{}as (IGFAE), Universidade de Santiago de Compostela, Santiago de Compostela, Spain
\item[$^{77}$] IMAPP, Radboud University Nijmegen, Nijmegen, The Netherlands
\item[$^{78}$] Nationaal Instituut voor Kernfysica en Hoge Energie Fysica (NIKHEF), Science Park, Amsterdam, The Netherlands
\item[$^{79}$] Stichting Astronomisch Onderzoek in Nederland (ASTRON), Dwingeloo, The Netherlands
\item[$^{80}$] Universiteit van Amsterdam, Faculty of Science, Amsterdam, The Netherlands
\item[$^{81}$] Case Western Reserve University, Cleveland, OH, USA
\item[$^{82}$] Colorado School of Mines, Golden, CO, USA
\item[$^{83}$] Department of Physics and Astronomy, Lehman College, City University of New York, Bronx, NY, USA
\item[$^{84}$] Michigan Technological University, Houghton, MI, USA
\item[$^{85}$] New York University, New York, NY, USA
\item[$^{86}$] University of Chicago, Enrico Fermi Institute, Chicago, IL, USA
\item[$^{87}$] University of Delaware, Department of Physics and Astronomy, Bartol Research Institute, Newark, DE, USA
\item[] -----
\item[$^{a}$] Max-Planck-Institut f\"ur Radioastronomie, Bonn, Germany
\item[$^{b}$] also at Kapteyn Institute, University of Groningen, Groningen, The Netherlands
\item[$^{c}$] School of Physics and Astronomy, University of Leeds, Leeds, United Kingdom
\item[$^{d}$] Fermi National Accelerator Laboratory, Fermilab, Batavia, IL, USA
\item[$^{e}$] Pennsylvania State University, University Park, PA, USA
\item[$^{f}$] Colorado State University, Fort Collins, CO, USA
\item[$^{g}$] Louisiana State University, Baton Rouge, LA, USA
\item[$^{h}$] now at Graduate School of Science, Osaka Metropolitan University, Osaka, Japan
\item[$^{i}$] Institut universitaire de France (IUF), France
\item[$^{j}$] now at Technische Universit\"at Dortmund and Ruhr-Universit\"at Bochum, Dortmund and Bochum, Germany
\end{description}

\section*{Acknowledgments}

\begin{sloppypar}
The successful installation, commissioning, and operation of the Pierre
Auger Observatory would not have been possible without the strong
commitment and effort from the technical and administrative staff in
Malarg\"ue. We are very grateful to the following agencies and
organizations for financial support:
\end{sloppypar}

\begin{sloppypar}
Argentina -- Comisi\'on Nacional de Energ\'\i{}a At\'omica; Agencia Nacional de
Promoci\'on Cient\'\i{}fica y Tecnol\'ogica (ANPCyT); Consejo Nacional de
Investigaciones Cient\'\i{}ficas y T\'ecnicas (CONICET); Gobierno de la
Provincia de Mendoza; Municipalidad de Malarg\"ue; NDM Holdings and Valle
Las Le\~nas; in gratitude for their continuing cooperation over land
access; Australia -- the Australian Research Council; Belgium -- Fonds
de la Recherche Scientifique (FNRS); Research Foundation Flanders (FWO),
Marie Curie Action of the European Union Grant No.~101107047; Brazil --
Conselho Nacional de Desenvolvimento Cient\'\i{}fico e Tecnol\'ogico (CNPq);
Financiadora de Estudos e Projetos (FINEP); Funda\c{c}\~ao de Amparo \`a
Pesquisa do Estado de Rio de Janeiro (FAPERJ); S\~ao Paulo Research
Foundation (FAPESP) Grants No.~2019/10151-2, No.~2010/07359-6 and
No.~1999/05404-3; Minist\'erio da Ci\^encia, Tecnologia, Inova\c{c}\~oes e
Comunica\c{c}\~oes (MCTIC); Czech Republic -- GACR 24-13049S, CAS LQ100102401,
MEYS LM2023032, CZ.02.1.01/0.0/0.0/16{\textunderscore}013/0001402,
CZ.02.1.01/0.0/0.0/18{\textunderscore}046/0016010 and
CZ.02.1.01/0.0/0.0/17{\textunderscore}049/0008422 and CZ.02.01.01/00/22{\textunderscore}008/0004632;
France -- Centre de Calcul IN2P3/CNRS; Centre National de la Recherche
Scientifique (CNRS); Conseil R\'egional Ile-de-France; D\'epartement
Physique Nucl\'eaire et Corpusculaire (PNC-IN2P3/CNRS); D\'epartement
Sciences de l'Univers (SDU-INSU/CNRS); Institut Lagrange de Paris (ILP)
Grant No.~LABEX ANR-10-LABX-63 within the Investissements d'Avenir
Programme Grant No.~ANR-11-IDEX-0004-02; Germany -- Bundesministerium
f\"ur Bildung und Forschung (BMBF); Deutsche Forschungsgemeinschaft (DFG);
Finanzministerium Baden-W\"urttemberg; Helmholtz Alliance for
Astroparticle Physics (HAP); Helmholtz-Gemeinschaft Deutscher
Forschungszentren (HGF); Ministerium f\"ur Kultur und Wissenschaft des
Landes Nordrhein-Westfalen; Ministerium f\"ur Wissenschaft, Forschung und
Kunst des Landes Baden-W\"urttemberg; Italy -- Istituto Nazionale di
Fisica Nucleare (INFN); Istituto Nazionale di Astrofisica (INAF);
Ministero dell'Universit\`a e della Ricerca (MUR); CETEMPS Center of
Excellence; Ministero degli Affari Esteri (MAE), ICSC Centro Nazionale
di Ricerca in High Performance Computing, Big Data and Quantum
Computing, funded by European Union NextGenerationEU, reference code
CN{\textunderscore}00000013; M\'exico -- Consejo Nacional de Ciencia y Tecnolog\'\i{}a
(CONACYT) No.~167733; Universidad Nacional Aut\'onoma de M\'exico (UNAM);
PAPIIT DGAPA-UNAM; The Netherlands -- Ministry of Education, Culture and
Science; Netherlands Organisation for Scientific Research (NWO); Dutch
national e-infrastructure with the support of SURF Cooperative; Poland
-- Ministry of Education and Science, grants No.~DIR/WK/2018/11 and
2022/WK/12; National Science Centre, grants No.~2016/22/M/ST9/00198,
2016/23/B/ST9/01635, 2020/39/B/ST9/01398, and 2022/45/B/ST9/02163;
Portugal -- Portuguese national funds and FEDER funds within Programa
Operacional Factores de Competitividade through Funda\c{c}\~ao para a Ci\^encia
e a Tecnologia (COMPETE); Romania -- Ministry of Research, Innovation
and Digitization, CNCS-UEFISCDI, contract no.~30N/2023 under Romanian
National Core Program LAPLAS VII, grant no.~PN 23 21 01 02 and project
number PN-III-P1-1.1-TE-2021-0924/TE57/2022, within PNCDI III; Slovenia
-- Slovenian Research Agency, grants P1-0031, P1-0385, I0-0033, N1-0111;
Spain -- Ministerio de Ciencia e Innovaci\'on/Agencia Estatal de
Investigaci\'on (PID2019-105544GB-I00, PID2022-140510NB-I00 and
RYC2019-027017-I), Xunta de Galicia (CIGUS Network of Research Centers,
Consolidaci\'on 2021 GRC GI-2033, ED431C-2021/22 and ED431F-2022/15),
Junta de Andaluc\'\i{}a (SOMM17/6104/UGR and P18-FR-4314), and the European
Union (Marie Sklodowska-Curie 101065027 and ERDF); USA -- Department of
Energy, Contracts No.~DE-AC02-07CH11359, No.~DE-FR02-04ER41300,
No.~DE-FG02-99ER41107 and No.~DE-SC0011689; National Science Foundation,
Grant No.~0450696, and NSF-2013199; The Grainger Foundation; Marie
Curie-IRSES/EPLANET; European Particle Physics Latin American Network;
and UNESCO.
\end{sloppypar}

}


\begin{thebibliography}{99}

\bibitem{auger_nim}
A.~Aab \textit{et al.} [Pierre Auger Collaboration],
Nucl.\ Instrum.\ Meth.\ A \textbf{798} (2015), 172--213,
\href{https://doi.org/10.1016/j.nima.2015.06.058}{doi:10.1016/j.nima.2015.06.058}.

\bibitem{uhecr24_overview}
D.~Boncioli \textit{et al.} [Pierre Auger Collaboration],
PoS(UHECR2024)027,
\href{https://doi.org/10.22323/1.484.0027}{doi:10.22323/1.484.0027}.

\bibitem{auger_prime}
A.~Castellina \textit{et al.} [Pierre Auger Collaboration],
EPJ Web Conf.\ \textbf{210} (2019), 06002,
\href{https://doi.org/10.1051/epjconf/201921006002}{doi:10.1051/epjconf/201921006002}.

\bibitem{auger_uub}
A.~Abdul Halim \textit{et al.},
JINST \textbf{18} (2023) P10016,
\href{https://doi.org/10.1088/1748-0221/18/10/P10016}{doi:10.1088/1748-0221/18/10/P10016}.

\bibitem{buscemi-jinst}
M.~Buscemi \textit{et al.},
JINST \textbf{15} (2020) P07011,
\href{https://doi.org/10.1088/1748-0221/15/07/P07011}{doi:10.1088/1748-0221/15/07/P07011}.

\bibitem{spmt_proceeding}
G.~A.~Anastasi \textit{et al.} [Pierre Auger Collaboration],
PoS(ICRC2023)343,
\href{https://doi.org/10.22323/1.444.0343}{doi:10.22323/1.444.0343}.

\bibitem{uub_proceeding}
M.~Bohacova \textit{et al.} [Pierre Auger Collaboration],
this proceedings.

\bibitem{ssd_calibration}
P.~Filip \textit{et al.} [Pierre Auger Collaboration],
PoS(UHECR2024)085,
\href{https://doi.org/10.22323/1.484.0085}{doi:10.22323/1.484.0085}.

\bibitem{ssd_performance}
M.~Conte \textit{et al.} [Pierre Auger Collaboration],
this proceedings.

\bibitem{ref:ICRC2025RDFirstData}
B.~Pont \textit{et al.} [Pierre Auger Collaboration],
this proceedings.

\bibitem{ref:ARENA2021Tomas}
T.~Fodran \textit{et al.} [Pierre Auger Collaboration],
PoS(ARENA2022)425,
\href{https://doi.org/10.22323/1.424.0043}{doi:10.22323/1.424.0043}.

\bibitem{icrc23_umd}
J.~de Jesús \textit{et al.} [Pierre Auger Collaboration],
PoS(ICRC2023)267,
\href{https://doi.org/10.22323/1.444.0267}{doi:10.22323/1.444.0267}.

\bibitem{uhecr24_umd}
J.~de Jesús \textit{et al.} [Pierre Auger Collaboration],
PoS(UHECR2024)077,
\href{https://doi.org/10.22323/1.484.0077}{doi:10.22323/1.484.0077}.

\end{thebibliography}
\end{document}